\documentclass[%
 reprint,
superscriptaddress,
 amsmath,amssymb,
 aps,
]{revtex4-1}

\usepackage{graphicx}
\usepackage{dcolumn}
\usepackage{bm}

\begin{document}

\preprint{APS/123-QED}

\title{Noise, delocalization and quantum diffusion in 1D tight binding model}

\author{Ehsan Gholami}
 \email{egholami@alum.sharif.edu}
\affiliation{Department of Physics, Isfahan University of Technology, Isfahan , Iran}
\author{Zahra Mohammaddoust Lashkami}%
\affiliation{Department of Physics, Zanjan University, Zanjan, Iran }
\date{\today}

\begin{abstract}
As an unusual type of anomalous diffusion behavior, the (transient) superballistic transport has been experimentally observed recently but it is not well understood yet. In this paper, we investigate the white noise effect (in Markov approximation) on the quantum diffusion in 1D tight-binding model with periodic, disordered and quasi-periodic region of size L attached to two perfect lattices at both ends in which the wave packet is initially located at the center of the sublattice. We find that in some completely localized system inducing noise could delocalize the system to desirable diffusion phase. This controllable system may be used to investigate the interplay of disorder and white noise, as well as exploring exotic quantum phase.
\begin{description}
\item[PACS numbers]
05.60.Gg, 03.65.Ϫw, 72.20.Dp, 05.45.Mt, 05.40.Ca
\end{description}
\end{abstract}

\pacs{Valid PACS appear here}
\maketitle


\section{\label{sec:level1}Introduction}

The quantum diffusion in 1D tight-binding model has an enriched background \cite{PhysRevA.36.5349,PhysRevE.55.4951,PhysRevE.90.042115,2016arXiv160304014G} . First, a numerical evidence was constructed by L. Hufnagel et al  \cite{PhysRevE.64.012301} supporting that the variance of a wave packet in 1D tight-binding can show a superballistic increase  (${\sigma^{2} = t^\nu}$ with $2 < \nu \leqslant 3$) for parametrically large time intervals with the appropriate model. 
They replaced the disordered part by a point source in which anything emitted from it could move
with a constant velocity modeling the dynamics of a perfect lattice\cite{PhysRevE.64.012301}.The model explains this phenomenon and its predictions were verified numerically for various periodic, disordered and quasiperiodic systems. Then, the superballistic diffusion of entanglement was constructed in disordered spin chains \cite{PhysRevA.72.050301}. In 2012 Z. Zhang et al \cite{PhysRevLett.108.070603} found a superballistic increase in variance ${\sigma^{2} = t^\nu}$ with $3 < \nu \leqslant 4.7$ numerically and extended the interpretation given in Ref. \cite{PhysRevE.64.012301} to diffusion rates beyond cubic. The superballistic growth of the variance has been experimentally observed for optical wave packets in 2013 for the first time \cite{Stutzer:13}.\\
The fractal \cite{PhysRevA.36.5349,Gmachowski2013194,PhysRevLett.108.093002} and multifractal \cite{PhysRevLett.79.1959} analysis of the width of a spreading wave packet revealed that for systems
where the shape of the wave packet is preserved, the k-th moment evolves as $t^{k\beta}$  with $\beta=D_{2}^{\mu}/D_{2}^{\Psi}$,where in general, $t^{k\beta}$ is an optimal lower bound, $D_{2}^{\mu}$ is the correlation dimension of the spectral measure $\mu$ (i.e. the local density of states) and $D_{2}^{\Psi}$ is the correlation dimension of the (suitably averaged) eigenfunctions.
The (disorder or phase-averaged) diffusion exponent is of particular physical importance because it characterizes the low-temperature behavior of the direct conductivity as given by Kubo's formula in the relaxation time
approximation\cite{Bellissard2000}.\\
Here in this article, we describe the system in universal terms, not specific to matter waves, as manifested by the analogy between the Schrodinger equation and the paraxial wave equation. Hyper-diffusion is in fact a universal concept, which should be observable in a variety
of systems beyond matter waves, such as optics, sound waves, plasma, as well as conducting electrons transport in semiconductors \cite{Levi2012}. Furthermore, fundamentally, once such temporal acceleration causes reaching high velocities, the relativistic effects have to be included.
These ideas have opened a range of exciting possibilities. 
However, in view of the recent experiments on quantum walks of correlated photons \cite{Peruzzo1500} and on localization with entangled photons \cite{PhysRevA.86.040302},it is very interesting to know whether the phenomenon of hyper-transport occurs with entangled photons as it occurs in entangled spin chains \cite{PhysRevA.72.050301}.\\

Taking the decoherence problem into account, the temperature effect on the spreading of wave packet is an essential feature \cite{PhysRevE.87.063104}.
The theoretical description of relaxation and decoherence processes in open quantum systems often leads to a non-Markovian dynamics which is determined by pronounced memory effects. Strong system-environment couplings, correlations and entanglement in the initial state, interactions with environments at low temperatures and with spin baths, finite reservoirs, and transport processes in nanostructures can lead to long memory times and to a failure of the Markovian approximation \citep{PhysRevA.75.022103}. But since here we intend to investigate the effect of white noise on superballistic diffusion in 1D tight-binding lattice, we do not deal with these restrictions, we can use the Markov approximation and the Lindblad equation. In Strong system-environment couplings we can also use a similar method called Non-Markovian generalization of the Lindblad theory of open quantum systems \citep{PhysRevA.75.022103,PhysRevLett.112.113601}.\\
In this work we examined the noise effect on the quantum wave packet dynamics in several nonuniform 1D tight-binding lattices, where a sublattice with on-site potential is embedded in a lattice with uniform potential. Irrespective of whether the sublattice on-site potential is periodic, disordered or quasiperiodic (some cases were studied in absence of any environment in Ref. \cite{PhysRevE.64.012301,PhysRevLett.108.070603}). We have found the threshold values of the white noise strength, beyond which the quantum superballistic diffusion does not occur (in the disordered case, the observed disappearance of superballistic diffusion is based on a fixed number of realizations of the sublattice). Such threshold values for disappearance of quantum superballistic diffusion should be one key element in real experimental studies, where the environment and noise have significant dephasing effect. Furthermore, based on our numerical studies we predict that the quantum diffusion exponent can be extensively tuned via the amount of induced white noise. Thus, we can manually induce noise to a system to drive it to a desired diffusion rate. The results must be within reach of today’s cold-atom experiments. 
\section{\label{sec:level2}The System}
\subsection{\label{sec:level2}The Lattice}
In this work we examine quantum wave packet dynamics in several nonuniform 1D tight-binding lattices, where a sublattice with on-site potential is embedded in a uniform lattice without on-site potential. Assume the following 1D tight-binding Hamiltonian
\begin{equation}\label{Ham}
H=-\Sigma_{i,j}(t_{ij}c^{\dagger}_{i}c_{j}+t^{*}_{ij}c^{\dagger}_{j}c_{i})+\Sigma_{i}V_{i}c^{\dagger}_{i}c_{i}
\end{equation}

where $t_{ij}$ is the tunneling rate from site i to site j and we set  $t_{ij}=t^{*}_{ij}=-1$ where $i=j\pm1$ and $t_{ij}=t^{*}_{ij}=0$ elsewhere, $c^{\dagger}_{i}$ and $c_{j}$ are the usual creation and annihilation operators, and $V_{i}$ represents the dimensionless on-site potential scaled by a tunneling rate, the sublattice have length $2L+1$, and it is located at the center of the perfect lattice.\\
The on-site potential is as follows:
\[
    V_{i}=\left\{
                \begin{array}{ll}
                  0&if \quad r_{i}\notin[-L,L] \\
                  W_{i}
                  
                \end{array}
              \right.
  \]
where $W_{i}$ can be periodic, semi-periodic  or disordered type.
\begin{figure}[h]
\includegraphics[width=.96\linewidth]{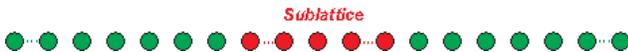}
  \caption{\begin{footnotesize}
  The schematic of the 1D lattice, the red circles (central part) represent sites with on-site potential while the green ones (circles in the left and right part) represent the sites without any on-site potential.
  \end{footnotesize} }
\end{figure}
 \\
At time zero a localized wave packet was launched in the sublattice's center, with $\rho_{m,n}=\delta_{c,c}$ where "c" denotes the central cite in the lattice. This initial state was a coherent superposition of many quasimomentum eigenstates. We have used the master equation of the Lindblad form that can be written in the form: 
\begin{equation}\label{lindblad} \begin{split} & \dfrac{\partial\rho}{\partial t}=L\rho =\\ & -\dfrac{i}{\hslash }[H,\rho]+\sum_{i} \dfrac{1}{2}\gamma_{i}M_{i}(2A_{i}\rho A_{i}^{\dagger}-\rho A_{i}^{\dagger} A_{i}-  A_{i}^{\dagger} A_{i}\rho)\\ & +\sum_{i} \dfrac{1}{2}\gamma_{i}N_{i}(2A_{i}\rho A_{i}^{\dagger}-\rho A_{i}^{\dagger} A_{i}-  A_{i}^{\dagger} A_{i}\rho) \end{split} \end{equation} \\where $H$ is Hermitian, $\rho$ is the density matrix $M_{i}$ and $N_{i}$ are real dimensionless non-negative c-numbers(either of which can be the larger), and the $A_{i}$ are arbitrary dimensionless operators.
The coefficient $\dfrac{1}{2}\gamma_{i}$ is a positive damping rate, and could in principle be absorbed into the other two coefficients, which would then acquire its dimension of inverse time.
If a master equation is of the Lindblad form, the time evolution is completely consistent with quantum mechanics, and in particular, this means that\\
i) The solution for the density operator is always a positive definite operator—
that is, no negative probabilities occur.\\
ii) The trace of $\rho$ is time independent, so that probability is conserved.
\\So the new Lindblad equation for white noise can be rewritten as: 
\begin{equation}\label{lindblad2}\dfrac{\partial\rho}{\partial t}=L\rho=-\dfrac{i}{\hslash }[H,\rho]+\Gamma\sum_{i} (2A_{i}\rho A_{i}^{\dagger}-\rho A_{i}^{\dagger} A_{i}-  A_{i}^{\dagger} A_{i}\rho) \end{equation}$ where \Gamma$ denotes the noise intensity. We assume that the interaction of the system with environment is dominated by white-noise captured within the Haken-Strobl model (pure-dephasing)\cite{Haken1973} . The dephasing term damps the all off-diagonal entries of the density matrix via the generators $A_{i} =|i\rangle \langle i | $, suppressing superposition
 of localized states at a rate $ \Gamma_{i}$, which is called the dephasing rate. Note that the pure-dephasing (Haken-Strobl) model is a simplified but useful model that has been successfully used in numerous studies in quantum optics, quantum information science, physical chemistry, and condensed matter physics. Its prediction becomes more realistic when the system is interacting with a thermal bath at high temperatures, where its effects can be modeled by white noise \cite{Novo2016} .
\subsection{\label{sec:level2}Measuring the Spreading of the Wave Packet}
We measure the spreading of the wave packet by its variance. The variance of the wave packet is defined as:\begin{equation}\label{variance}
variance\equiv\sigma^{2}\equiv\sum_{n}n^{2}{|\psi_{n}|}^2
\end{equation}
where n is the lattice site index and $\psi_{n}$depicts a normalized time-evolving wave packet. 

\begin{figure}[h]
\includegraphics[width=.96\linewidth]{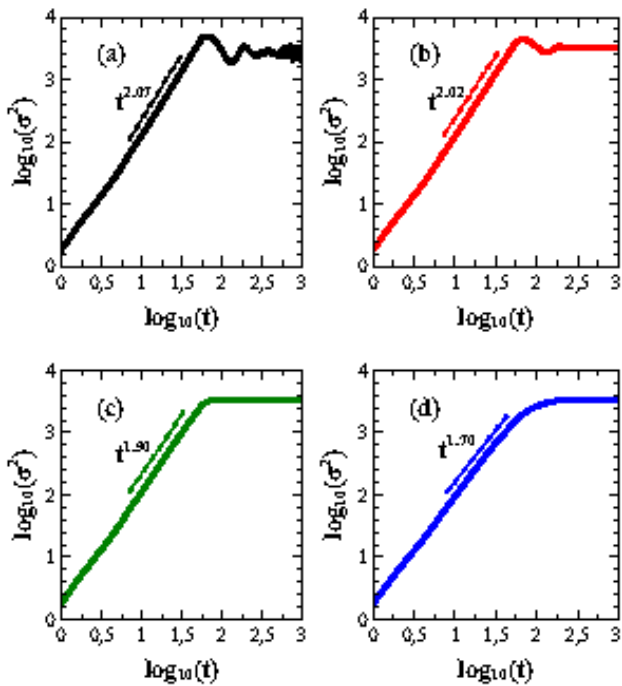}
  \caption{\begin{footnotesize}
Time dependency of the variance of the wave packet (${\sigma}^2$ ) for a periodic potential with a sublattice potential intensity, $V=0.5$ (with L = 10, $2L+1$ is  the sublattice's size). From top-left to bottom-right for noise intensity $\Gamma=0,0.01,0.04, $ and $ 0.1$ respectively, here and in all other figures, the dashed lines represent power-law fitting. The all quantities are dimensionless.
  \end{footnotesize} }
\end{figure}
\begin{figure}[h]
\includegraphics[width=.96\linewidth]{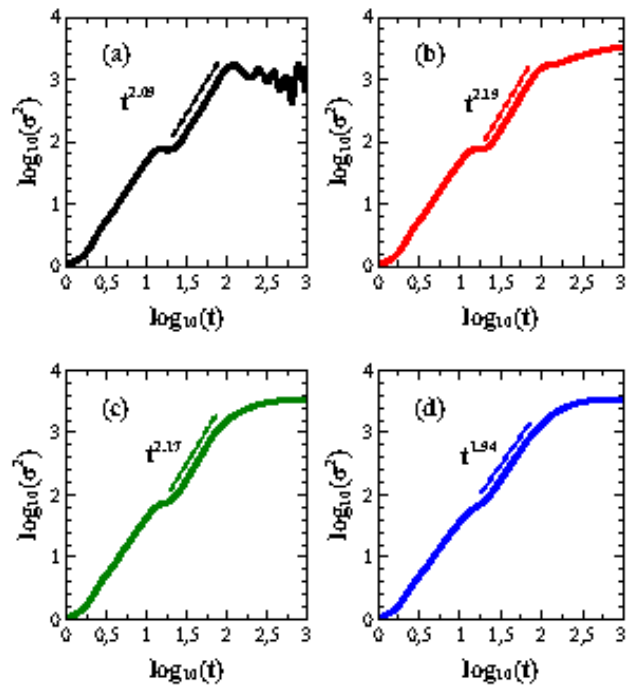}
  \caption{\begin{footnotesize}Same as Figure 2, for a periodic potential with a sublattice potential intensity, $V=1.5$.
  \end{footnotesize} }
\end{figure}
\section{\label{sec:level2}Computation Cases}

\subsection{\label{sec:level2}Periodic Case}
Here we chose a periodic on-site potential in the sublattice i.e. we let $W_{i}=0.5{(-1)^i}$. The time dependence of the variance of the wave packet ${\sigma}^2(t)$ for sublattice potential intensity $V=0.5$ is shown in Fig. 2. The spreading of the wave packet in absence of any noise is shown in Fig. 2(a). We study the effect of the white noise with small noise intensity $\Gamma=0.01$ in Fig.2(b). As it can be seen the white noise reduces the diffusion exponent by amount of 0.05. If we increase the noise intensity to $\Gamma=0.04$ (see Fig. 2(c)) the white noise reduces the diffusion exponent to amount 1.90 this means that in this case the diffusion rate is changed from superballistic regime to subballistic. Further increasing the noise intensity to $\Gamma=0.1$ (as you can see in Fig. 2(d)) can decrease the diffusion exponent by amount of 0.37 with respect to the noise free case.
It is seen that the white noise suppresses the diffusion significantly for all the above cases. But if we increase the sublattice potential intensity to $V=1.5$, the white noise affects the diffusion differently. In absence of any noise when we set the sublattice potential intensity to $V=1.5$, it appears that the diffusion exponent increases to 2.09 (seen Fig. 3(a)). The effect of white noise with small noise intensity $\Gamma=0.01$ changes dramatically from what we have seen in Fig. 2(b). We see that the white noise shows a counter-intuitive effect on the diffusion rate. Here noise is not a nuisance to be avoided anymore but it improves the diffusion exponent by amount of 0.10 (see Fig .3(b)). This bizarre behavior is due to the fact that the white noise has delocalized some of the system states. More importantly, the delocalized states are found to separate energy domains corresponding to two distinct types of localized states: the usual localized states centered at sites in the sublattice and a new type of states which are localized at the contact with the perfect lattice, which were named antilocalized states \cite{PhysRevB.51.5699}.
The antilocalized states are expected to play a special role in improving the diffusion rate.
When noise intensity $\Gamma$ increases from 0.01 to 0.04(see Fig .3(c)) the diffusion exponent decreases a little (0.02), here the stochastic resonance between the matter-wave antilocalized states and the white noise becomes less than what we see in Fig .3(b). So the system's diffusion gets a little slower with respect to the case with smaller noise intensity. Increasing the noise intensity to $\Gamma=0.1$, makes the stochastic resonance between the system's antilocalized states and the white noise disappear. As it can be seen in  Fig .3(d) this decreases the diffusion exponent more and makes a regime change in the system's diffusion from superballistic to subballistic.

\begin{figure}[h]
\includegraphics[width=.96\linewidth]{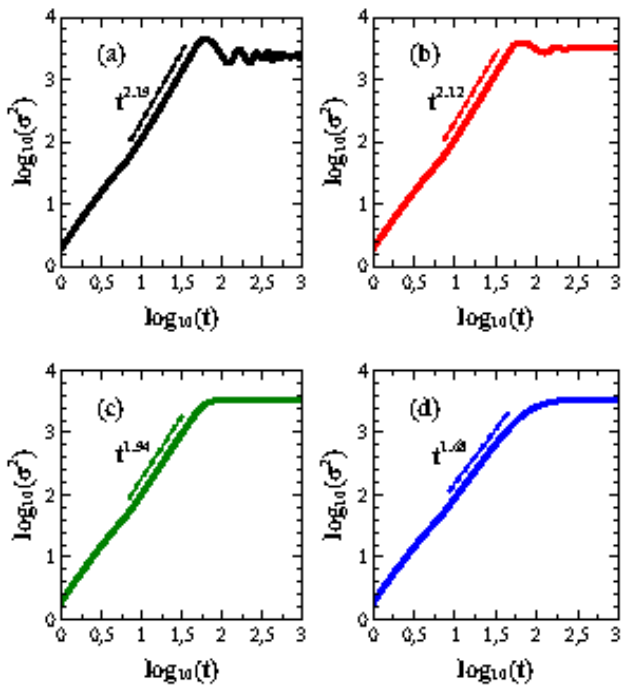}
  \caption{\begin{footnotesize}
Same as Figure 2, for a disordered sublattice with a sublattice potential intensity, $V=0.5$. 
  \end{footnotesize} }
\end{figure}
\subsection{\label{sec:level2}Disordered Case}
Here we choose a disordered on-site potential in the sublattice i.e. we let the $W_{i}$ takes $+V or -V$ randomly . Therefore, we obtain slightly different results of $\sigma^{2}(t)$ from different disorder realizations in [{-L}, L]. Finally, we present the result of $\sigma^{2}(t)$ after first averaging them over 50 different disorder realizations.
The time dependence of the averaged variance of the wave packet ${\sigma}^2(t)$ for sublattice potential intensity $V=0.5$ is shown in Fig. 4. The spreading of the wave packet in absence of any noise is shown in Fig. 4(a). We also study the effect of white noise with small noise intensity $\Gamma=0.01$, and it can be seen from Fig.4 (b) that the white noise reduces the diffusion exponent by 0.07. If we increase the noise intensity to $\Gamma=0.04$ (see Fig. 4(c)) the white noise reduces the diffusion exponent to 1.94. This means that in this case the diffusion rate is changed from superballistic regime to subballistic. Further increase in the noise intensity to $\Gamma=0.1$ (as you can see in Fig. 4(d)) can decrease the diffusion exponent by 0.51 with respect to the noise free case.\\

In absence of any noise when we set the sublattice potential intensity to $V=0.8$, it appears that the diffusion exponent increases to 2.33 (seen Fig. 5(a)). If we induce small noise intensity $\Gamma=0.01$ it decreases the diffusion exponent by 0.07 (see Fig .5(b)). 
When noise intensity $\Gamma$ increases from 0.01 to 0.04 the diffusion exponent decreases more (see Fig .5(c)). Increasing the noise intensity to $\Gamma=0.1$ reduces the diffusion exponent to 1.80 (see Fig .5(d). This means that in this case the diffusion changes from superballistic (hyper-diffusion) diffusion to subballistic diffusion.
Because in disordered sublattice there is higher intrinsic disorder with respect to the periodic case, there is no stochastic resonance between the system's antilocalized states and the white noise. So inducing white noise to the system cannot reduce the total noise of the system, so the white noise has its usual suppression role here.

\begin{figure}[h]
\includegraphics[width=.96\linewidth]{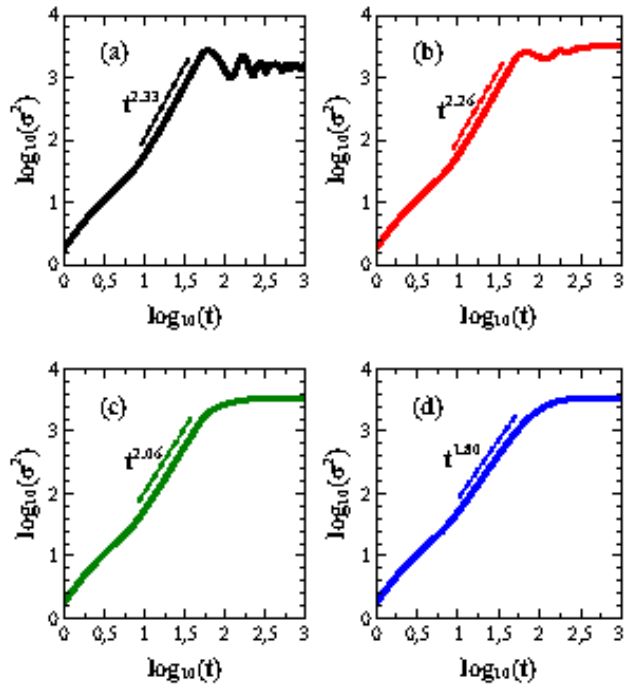}
  \caption{\begin{footnotesize}
Same as Figure 2, for a disordered sublattice with a sublattice potential intensity, $V=0.8$.
  \end{footnotesize} }
\end{figure}

\subsection{\label{sec:level2}Quasiperiodic case}
A quasiperiodic potential is intermediate between a truly random potential which may cause localization in one-dimensional (1D) systems and periodic potentials which leads to energy bands and extended states\cite{Pandit1873}. Quasiperiodic systems possess intriguing energy spectra and eigenstate structure\citep{fib1,fib2,fib3}. 

\subsubsection{\label{sec:level3}Fibonacci Case}

The other sublattice's potential that we will consider here is the simplest model of a quasicrystal, called the diagonal model, obtained when $W_{i}$ has two possible values, that we denote by $W_{A}$ and $W_{B}$, following the Fibonacci sequence
(FS). The FS is built as follows: consider two letters, A and B, and the substitution rules, $A \rightarrow B$; and $B \rightarrow AB$: if one defines the first generation sequence as $F_{1}=A$ and the second one as $F_{2}=B$, the subsequent chains are generated
using the two previous rules. For instance, $F_{3} = AB$: starting with an A, we construct the following sequences,
A, B, AB, BAB, ABBAB, BABABBAB, and so on. Each generation obtained by iteration of the rules is labeled with an index l: The number of letters in each generation l is given by the Fibonacci numbers F(l) of generation l, which satisfy: $F(l)=F(l-1)+F(l-2)$ with the initial conditions:
$F(0)=1; F(1)=1$. It is well known that a Fibonacci lattice yields singular continuous energy spectra and critical eigenstates (that are neither localized nor extended)\cite{fib1,fib2,fib3,fib4}. 

Here the sublattice has 55 sites and we choose sublattice Fibonacci potential with intensity $V=0.5$. If we set $W_{A}=+0.5$ and $W_{B}=-0.5$  in absence of any noise, it appears that the diffusion exponent increases to a high amount of 2.79 (see Fig. 6(a)). If we induce a small noise intensity $\Gamma=0.01$ it decreases the diffusion exponent by 0.24 (see Fig .6(b)).
When noise intensity $\Gamma$ increases from 0.01 to 0.04 the diffusion exponent decreases more (see Fig .6(c)). Increasing the noise intensity to $\Gamma=0.1$ reduces the diffusion exponent to 1.94 (see Fig .6(d)). This means that in this case the diffusion changes from superballistic diffusion (hyper-diffusion) to subballistic diffusion.  

In Fibonacci case the intrinsic disorder is higher than the periodic case but still smaller than the disordered case. Yet there is no stochastic resonance between the system's antilocalized states. Therefore the white noise suppresses the diffusion here. We see that the diffusion exponent can be tuned by changing $\Gamma$. 
\begin{figure}[h]
\includegraphics[width=.96\linewidth]{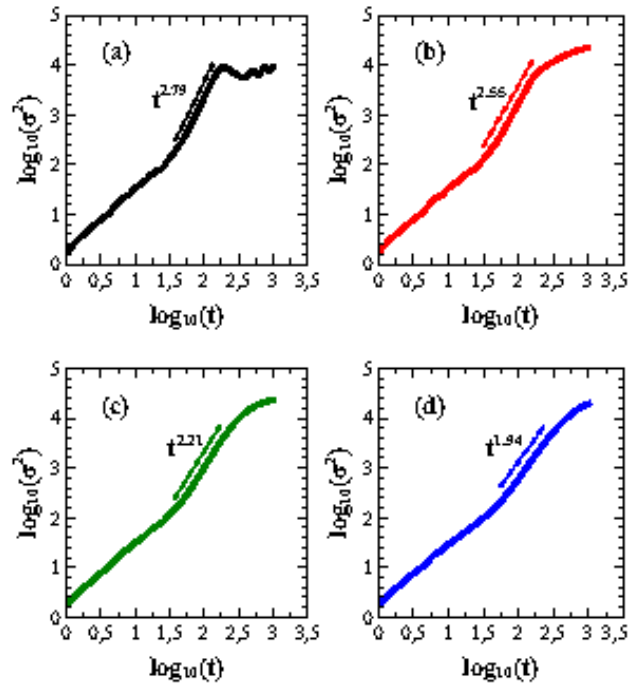}
  \caption{\begin{footnotesize}
Same as Figure 2, for a Fibonacci sublattice with a sublattice potential intensity, $V=0.5$ (with sublattice size {2L+1} = 55).  It is evident that the saturation point is much higher than cases with L = 10). 
  \end{footnotesize} }
\end{figure}

\subsubsection{\label{sec:level3}Harper Case}
We now turn to a non-interacting Harper sublattice\cite{Harper} with Aubry-Andre´ Hamiltonian\cite{Roati2008}.
\begin{equation}\label{Harper} \begin{split}
 H=&J\sum_{i}(|w_{i} \rangle\langle w_{i+1}|+|w_{i+1} \rangle\langle w_{i}|)+\\&{\Delta 
 \sum_{i\in[-L,L]}cos (2\pi\beta+\phi)|w_{i} \rangle\langle w_{i}|} \end{split} \end{equation}
where $|w_{i} \rangle$ is the Wannier state localized at the lattice site i, J is the site-to-site tunneling energy and $\Delta$ is the strength of the potential. 

\begin{figure}[h]
\includegraphics[width=.96\linewidth]{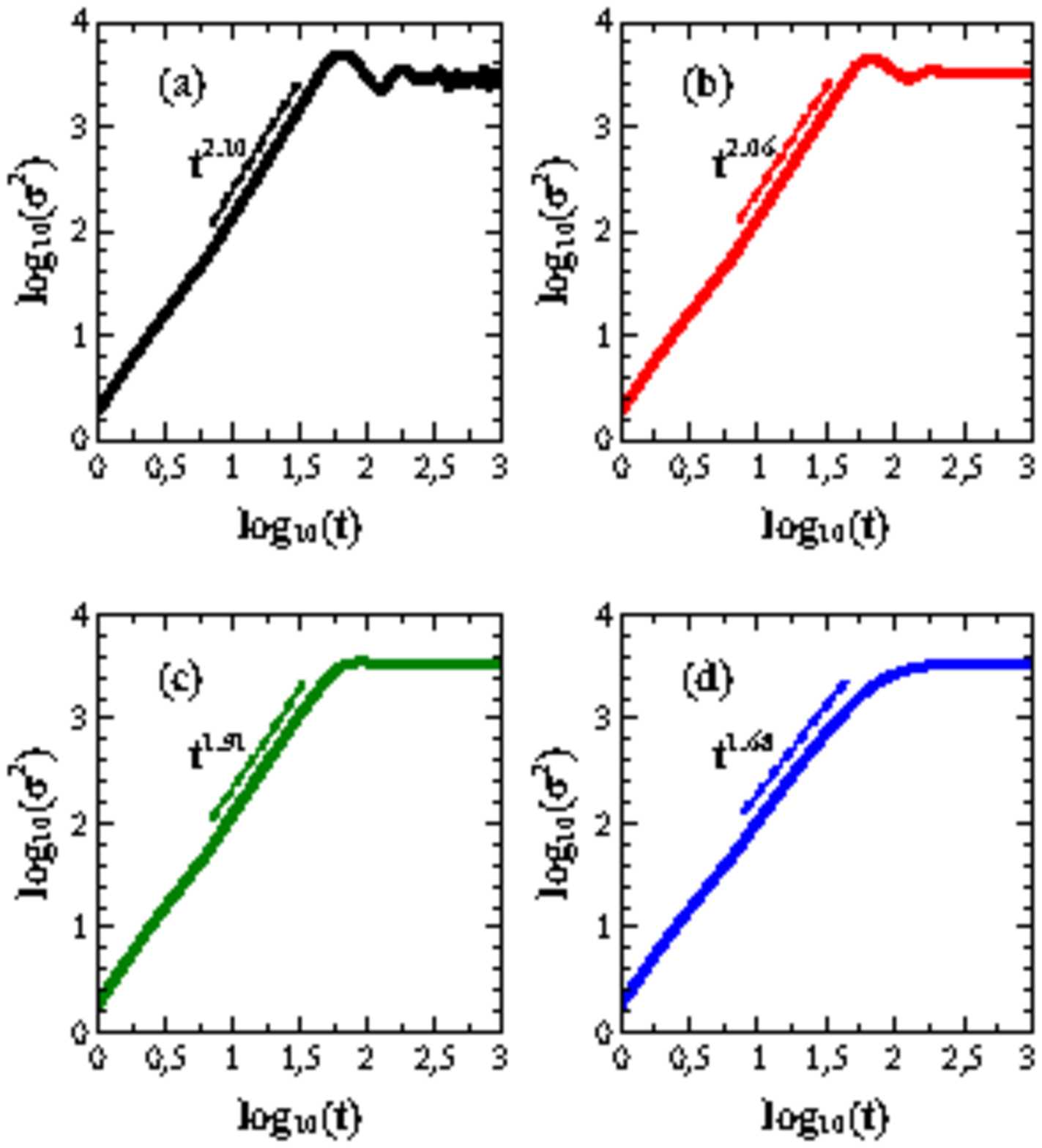}
  \caption{\begin{footnotesize}
Same as Figure 2, for a Harper sublattice with a sublattice potential with $\Delta/\j=0.5$. 
  \end{footnotesize} }
\end{figure}

\begin{figure}[h]
\includegraphics[width=.96\linewidth]{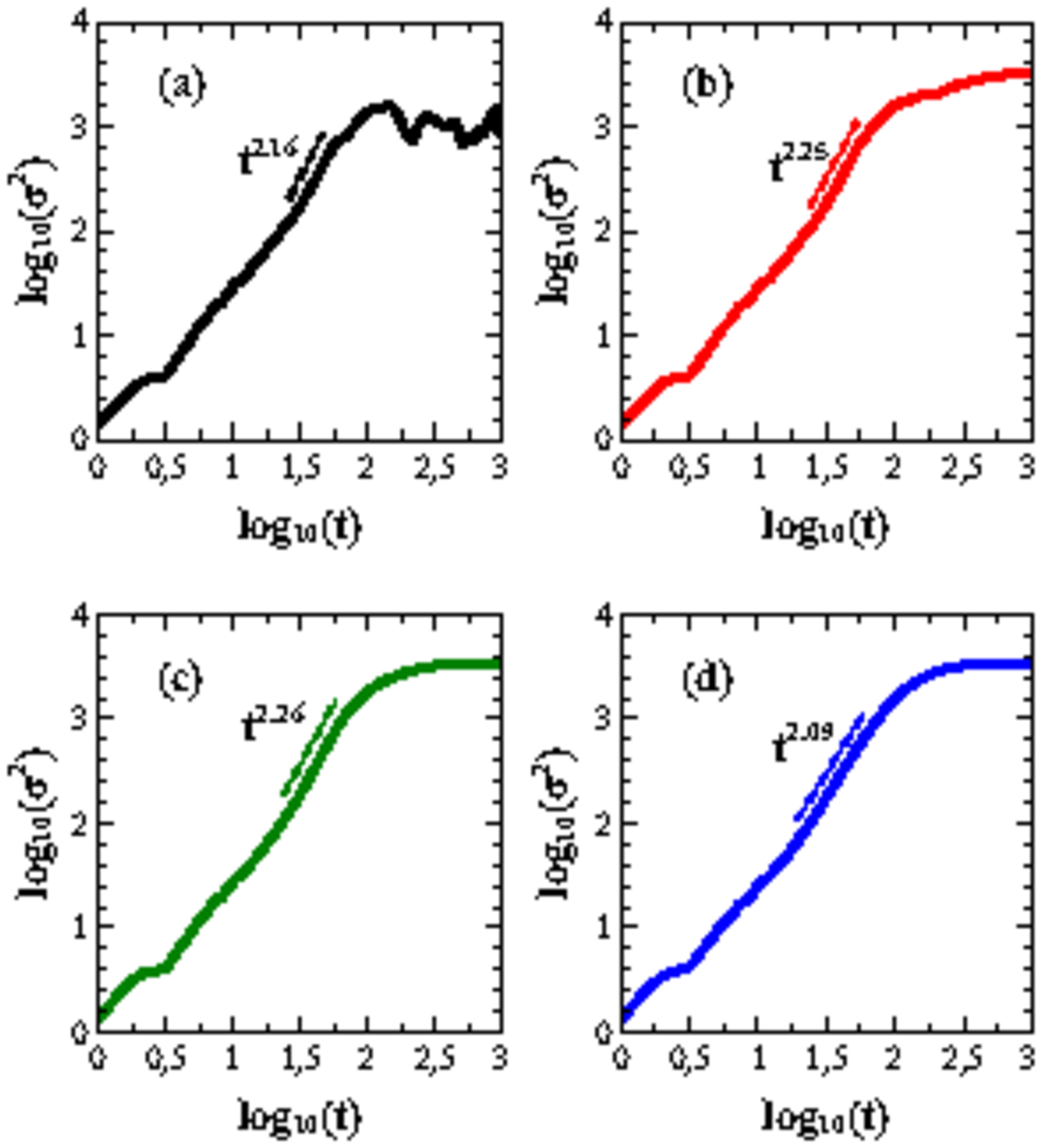}
  \caption{\begin{footnotesize}
Same as Figure 2, for a Harper sublattice with a sublattice potential with $\Delta/\j=1.5$. 
  \end{footnotesize} }
\end{figure}

\begin{figure}[h]
\includegraphics[width=.96\linewidth]{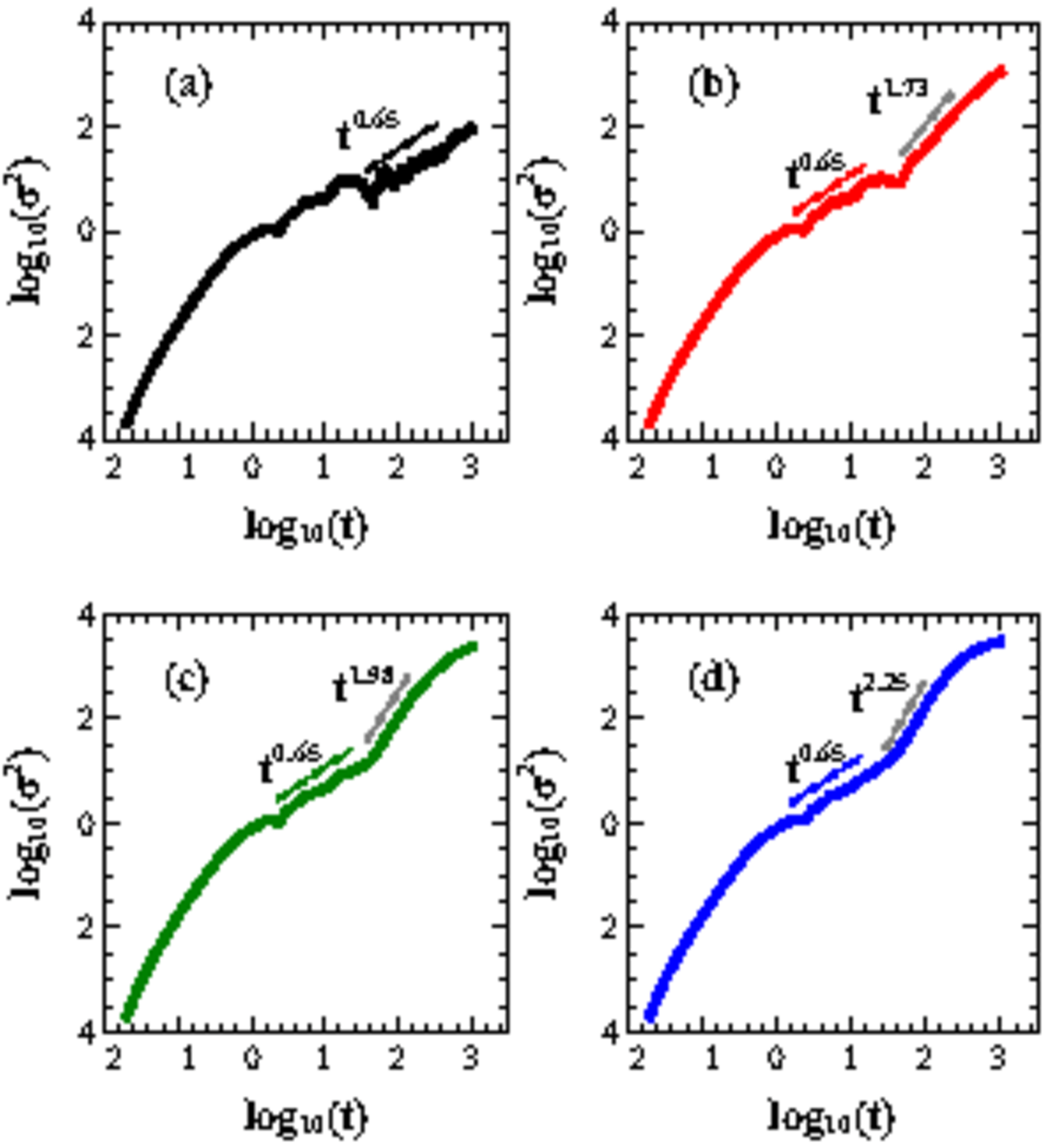}
  \caption{\begin{footnotesize}
Same as Figure 2, for a Harper sublattice with a sublattice potential with $\Delta/\j=2.5$. 
  \end{footnotesize} }
\end{figure}
$\beta={k_{2}}/{k_{1}}$ is the ratio of the two lattices wave numbers, and $\phi$ is an arbitrary phase (we set $\phi=0$).For a rational $\beta$, above equation can be solved by Bloch’s theorem. Although its value is very limited since the coefficients in Fourier space of the solution form a dense set. For $\beta$ irrational, the spectrum depends on the value of $\Delta$/$J$. 
 We have studied it for $\Delta/\j=0.5$ (see Fig. 7), $\Delta/\j=1.5$ (see Fig. 8), $\Delta/\j=2.5$ (see Fig. 9).
In the experiment, the two relevant energies $J$ and $\Delta$ can be controlled independently by changing the heights of the primary and secondary sublattice potentials respectively. For a maximally incommensurate ratio $\beta={({\sqrt{5}}-1)}/2$, the model exhibits a sharp transition from extended to localized states at ${\Delta}/{J}=2$\cite{Roati2008}. 
In Fig.7 we present the diffusion rate for Harper potential for $\Delta/\j=0.5$ in absence of any noise.  It turns out that the diffusion rate increases to $\sigma^{2}=t^{2.1}$ (see Fig. 7(a)). If we induce small noise intensity $\Gamma=0.01$ it decreases the diffusion exponent by 0.04 (see Fig .7(b)).
When noise intensity $\Gamma$ increases from 0.01 to 0.04 the white noise reduces the diffusion exponent to 1.91 (see Fig .7(c)). This means that in this case the diffusion changes from superballistic diffusion (hyper-diffusion) to subballistic. Increasing the noise intensity to $\Gamma=0.1$ (as you can see in Fig. 7(d)) can decrease the diffusion exponent by 0.42, with respect to the noise free case. 

In Harper case for $\Delta/\j=0.5$ the intrinsic disorder is higher than the periodic case, but smaller than the disordered and Fibonacci case. Hence there is no stochastic resonance between the system's antilocalized states and the white noise. Therefore the white noise suppresses the diffusion rate.
\\It is worthwhile mentioning that $\beta$ controls the transition from periodic to quasiperiodic sequences. Thus, the present approach also leads to the possibility of studying the electronic properties as a function of such parameter. Furthermore, it can be proven that the parameter $\beta$ can also be related with a magnetic field, as happens in the Harper potential. In that case, instead of having a constant magnetic field in space, one has a space modulated magnetic field \cite{opez-Rodr201406}. In real systems, the changes in $\beta$ are simple to study using many different devices, since
its effect is only a change in the sequence of the binary potential. For example, one can use microwaves in a cavity, a dielectric superlattice, or a space modulated magnetic field in a semiconductor.\cite{Naumis20081755} \\

When we set $\Delta/\j=1.5$, the white noise effects would be different. In absence of any noise, it appears that the diffusion exponent increases to 2.16 (see Fig. 8(a)). The effect of white noise with small noise intensity $\Gamma=0.01$ changes dramatically from what we have seen in in Fig .7(b), the white noise delocalizes some of the system states and shows it's counter-intuitive effect again with and improves the diffusion exponent (see Fig. 8(b)).
When noise intensity $\Gamma$ increases from 0.01 to 0.04(see Fig .8(c)) the diffusion exponent is still higher than the case in absence of the noise.

As it can be seen in Fig .8(d), increasing the noise intensity to $\Gamma=0.1$, reduces have some damping effect on the diffusion, makes the stochastic resonance between the system's antilocalized states and the white noise smaller.

For $\Delta/\j=2.5$ in absence of any noise, it appears that the diffusion exponent is as low as 0.65 (seen Fig. 9(a)). The white noise with small noise intensity $\Gamma=0.01$ does not change the first sub-diffusion region. But as time goes on, the white noise delocalizes some of the system states and drives system's sub-diffusion ($0<\nu<1$) to sub-ballistic diffusion ($1<\nu<2$) instantly (see Fig .9(b)).
When noise intensity $\Gamma$ increases from 0.01 to 0.04(see Fig .9(c)), first there is no change in the diffusion rate, but then the white noise delocalizes some of the system states and makes system to diffuse almost ballistically, here the stochastic resonance between the matter-wave antilocalized states and the white noise becomes more than what we see in Fig .9(b). So the system's diffusion gets faster compared to the smaller noise intensity. Increasing the noise intensity to $\Gamma=0.1$. Increasing the noise intensity to $\Gamma=0.1$, enhances the stochastic resonance between the system's antilocalized states and the white noise, such that it changes the system's diffusion from  subdiffusion to superballistic regime instantly(see Fig .9(d)).

\section{\label{sec:level1}Finite size effect}
Here, we study the variance of the wave packet for finite sized lattices. As it can be seen in the all figures the variance is saturated at a point which is different for different lattices. It is notable that in Fibonacci case with $W_{A}=+0.5$ and $W_{B}=-0.5$(see Fig. 6(a)) we find diffusion exponent 2.79 in absence of noise, while this diffusion exponent is considerably larger than what found in \cite{PhysRevLett.108.070603}. This means that sometimes the finite size could improve the diffusion rate significantly. This can be explained via the structure of the underlying eigenstates.
\begin{figure}[h]
\includegraphics[width=.96\linewidth]{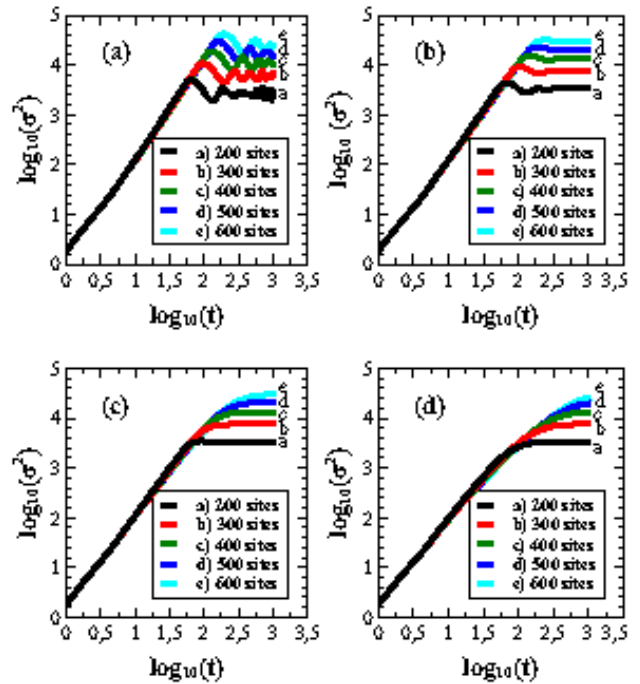}
  \caption{\begin{footnotesize}
Same as Figure 2, effect of finite size effect on periodic sublattices with different lattice sizes (200,300,400,500,600).
  \end{footnotesize} }
\end{figure}
There are many researches on the finite size effect in different physical systems\cite{hsci,PhysRevB.94.035126,1063-7869-47-6-R01,KRIVINE1986125,Guidi2015,4395255,Riboli:11,PhysRevB.53.1814} . The finite-size effect in the absence of coupling with the environment is manifested in Fig. 10(a)). As we increase the size of the lattice the final saturation point gets higher. Due to the practical limits we cannot solve the master equation for large lattices. But as we know from the Schrodinger equation's solution for much bigger lattices (in absence of any noise), this finite size effect is usually just in the point of saturation and the diffusion exponent. So our results will be different for other lattices and sub-lattices site numbers. But in the lab usually even a lattice with 200 sites is big enough for many experiments.
In calculation of the finite size effect in systems under white noise influence (see Fig. 10 (b-d)), we can see that as we increase the noise intensity, it damps the fluctuation more and more until there is no more fluctuation (see Fig. 10 (d)). As the lattice size increases from 200 sites to 600 sites in a periodic sub-lattice then the final saturation point becomes higher. We have studied other potentials as well in which the lattice size changes the saturation point in all cases. So, this long-time limit must be characterized by localization due to the finite size effect. The white noise does not change the total saturation point because it is a finite size effect.\\
\section{\label{sec:level1}Localization to delocalization transition}
The localization to delocalization transition has an old background \cite{{Dunlap},{Ishii01011973},{BRATUS1988449}}. Recently this subject get more attractive\cite{Bordenave2013,Eleuch}, Yamada have studied delocalization in one-dimensional tight-binding models with fractal disorder\cite{Yamada2015}, the one-dimensional tight-binding models with an ergodic and stationary random potential have positive Lyapunov exponent of the wavefunction with probability 1 (G-M-P theorem). A survey was made of some mathematical results and techniques for Schrödinger operators with random and quasiperiodic potentials in the ref.\cite{Spencer1988,Carmona1985}. The existence of the positive Lyapunov exponent is necessary and sufficient condition for a pure point set spectrum of the operators, and then all the eigenfunctions exhibit the exponentially decay in the thermodynamic limit. Kotani’s theory states that if the potential sequence is nondeterministic under the following conditions, (i) stationarity, (ii) ergodicity, (iii) integrability, then there is no absolutely continuous (a.c.) spectrum of the operators .These theorems can be proven true for continuous and discrete one-dimensional disordered systems (1DDS) \cite{Simon1984}. 
Usually in the presence of a background noise an increased effort put in controlling a system stabilizes its behavior. Rarely it is thought that an increased control of the system can lead to a looser response and, therefore, to a poorer performance\cite{PhysRevE.77.051107}.
For example here we consider a triangular potential with:\\ $W_{i}\in(W_{1}=V_{min},...,W_{N/2}=V_{max},...,V_{1}=W_{min})$ where N is the total sites number of sub-lattice (an odd number) and $V_{max}$($V_{min}$) is the biggest(smallest) on-site potential.

\begin{figure}[h]
\includegraphics[width=.96\linewidth]{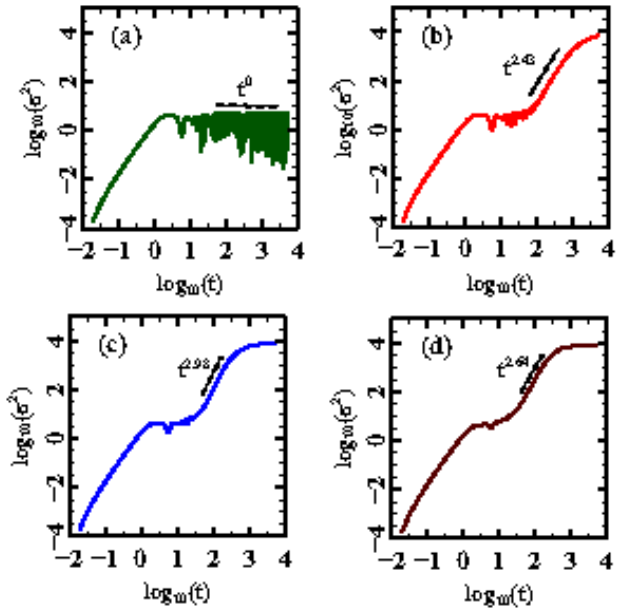}
  \caption{\begin{footnotesize}
Same as Figure 2, for a sublattice with a triangular potential (with L = 10 but larger perfect lattice) for more time steps.
  \end{footnotesize} }
\end{figure}

Keeping a quantum system in a given instantaneous eigenstate is a control problem\cite{2016arXiv160507523J} .
As it can be seen in Fig.11 (a) in absence of any noise (the increased control of the system) this Hamiltonian generates localization, and this is a manifestation of Anderson localization (see Fig. 11(a)), because as it is shown the saturation point is far beyond the amount that is dictated by finite size effect (see Fig. 11(b, c, d)). When we induce small noise intensity $\Gamma=0.01$ to the system (it make system to destabilize its behavior), it takes some times so the white noise delocalizes some of the system states and drives system to sub-diffusion very smoothly, then at some time step suddenly the stochastic resonance occurs and system changes it's diffusion regime instantly to superballistic regime (see Fig.11(b)). If we increase the noise intensity to $\Gamma=0.04$ (see Fig. 11(c)) the white noise delocalizing effect is similar to later case but stochastic resonance occurs earlier with respect to later case. Further increasing in the noise intensity to $\Gamma=0.1$ (as you can see in Fig. 11(d)) results  earlier stochastic resonance with respect to the former cases, as we could see here stochastic resonance can have more increasing effect on the diffusion rate at certain points. Further assessment of this stochastic resonance is presented in next section. Here the finite size and the stochastic resonance are correlated.

\section{\label{sec:level1}Signs of stochastic resonance}
The process whereby noise operates on the quantum system enhancing the response to an external noisy signal has been termed stochastic resonance.  Upon decreasing the temperature, quantum tunneling becomes increasingly important\cite{Grifoni.PhysRevLett.76.1611}.  Above a crossover noise intensity, noise activated transitions dominate over quantum tunneling events. The effects of quantum noise then result in a quantum correction factor of the classical rate of activation. As noise intensity is decreased below a threshold, tunneling transitions prevail. The quantum noise could be characterized by the temperature of the thermal bath and by the coupling strength of the system to the environment. In the absence of driving, at sufficiently high noise intensity the damping effects are so strong that quantum coherence is completely suppressed by incoherent tunneling transitions. We try to find the coherent tunneling transitions between adjacent sites in the border of the sublattice and perfect lattice. We define the coherent tunneling rate (QTR) as norm of the off-diagnal element of density matrix between two adjacent sites in the border of the sublattice and perfect lattice($\rho_{m,m+1}$). First we look at the coherent tunneling transitions (QTT) in system in which noise decreases diffusion rate. Here in absence of any noise, QTR is fluctuating in all time periods (see Fig .12(a)), as the noise increases, the fluctuation in QTR becomes damped sooner (see Fig .9(b,c,d)), by increasing the noise intensity the coherency will be killed very soon, this is the usual decoherence effect of the white noise. However in the systems that noise enhances the diffusion rate or delocalizes a completely localized system it has different role. In the triangular potential, increasing noise intensity has various effects on the QTT (see Fig. 13), in absence of any noise QTR is zero, as noise increases the pick of QTR shifted to later time steps, and the duration in which we have the coherency increases significantly. 
\begin{figure}[h]
\includegraphics[width=.96\linewidth]{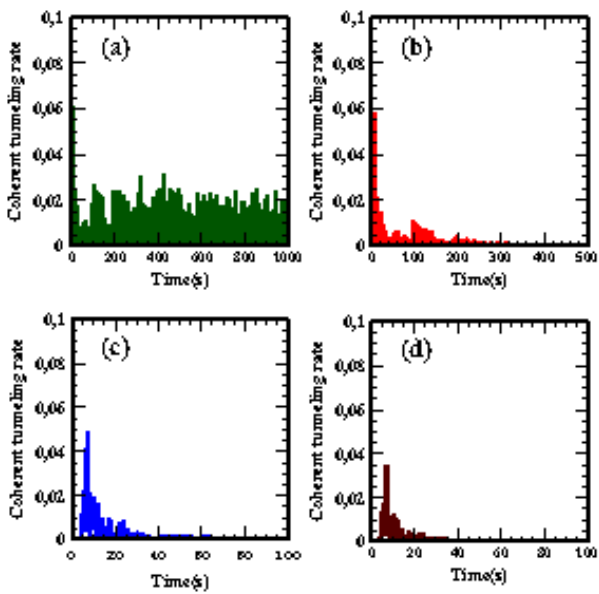}
  \caption{\begin{footnotesize}
Coherent tunneling rate between adjacent sites in the border of the sublattice and perfect lattice, for a constant potential with a sublattice potential intensity, $V=1.0$ (with L = 10). The plots from top-left to bottom-right belongs to cases with the noise intensity $\Gamma=0,0.01,0.04, $ and $ 0.1$ respectively. All quantities are dimensionless. 
  \end{footnotesize} }
\end{figure}

\begin{figure}[h]
\includegraphics[width=.96\linewidth]{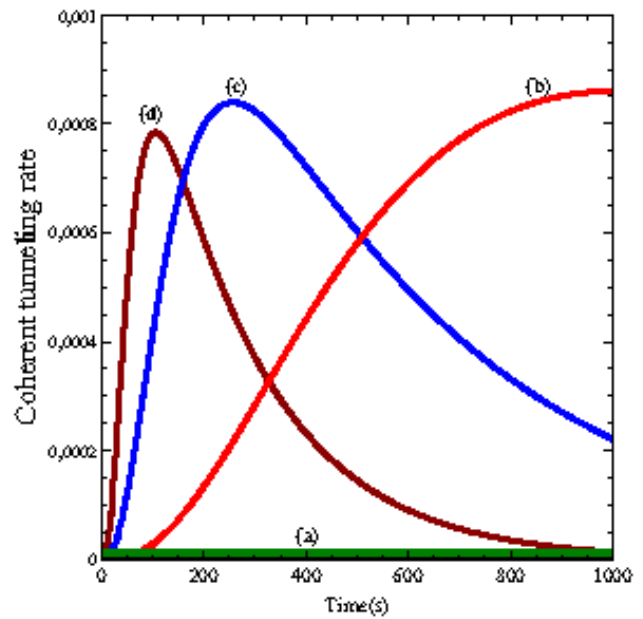}
  \caption{\begin{footnotesize}
Same as Figure 12, for a sublattice with a triangular potential (with L = 10 and larger perfect lattice).Green (a), red (b), blue (c) and the brown line (d) represents noise intensity $\Gamma=0,0.01,0.04, $ and $ 0.1$ respectively.
  \end{footnotesize} }
\end{figure}
\section{\label{sec:level1}Calculation method}
For a tight-binding Hamiltonian with N sites, the number
of elements in the density matrix is $N^2$, and solving the master
equation by numerical integration involves of
use of super-operators of size $N^2{\times}N^2$. we used sparse matrix format in which,
not all of the $N^4$ elements need to be stored in the memory.
However, the time required to evolve a quantum system according
to the master equation still increases rapidly as a function of the
system size, especially the amount of RAM that is needed becomes larger and larger. Consequently, solving the master equation is practical
only for relatively small systems: $N\lesssim1000$. We used Scipy \cite{scipy}(an open source Python library used for scientific computing and technical computing.) for integrating and sparsing the matrices. About the precision in the value of the exponents we should mention that in all cases the standard error of coefficients for the regression is less than $0.002$.
\section{\label{sec:level1}Discussion}
For explaining the appearance of
the superballistic diffusion, a simple probabilistic model
called the point-source model has been introduced in Ref.\cite{PhysRevE.64.012301} . The disordered sublattice can be replaced by a point source that radiates the
probability with a constant velocity v, simulating the dynamics
of the perfect lattice. At
$t = 0$, all probability is trapped inside the point source and
as time passes, the probability of finding the particle in the sublattice decays exponentially as ${P_{L}}(t)={exp}({-{\lambda} {t^2}})$, where $\lambda$  is the probability decay rate. 
 When t is small, Zhang et al argued that the exponential decay of the probability of finding the particle in the sublattice can be approximated as ${P_{L}}(t)={exp}({-{\lambda} {t^\alpha}})\approx 1-{\lambda} {t^\alpha}$, here $\gamma\approx\alpha+2$ \cite{PhysRevLett.108.070603}. If $\alpha=0$, there is no decay, If $\alpha=1$, then an
exact exponential decay occurs. In their own article they have mentioned a big difference between analytic calculation($\gamma\approx4.3$) and numerical one($\gamma\approx4.7$).\\ Recently Nguyen et al. have shown that the nonlinear fitting extension of the point-source model in the form that have introduced in Ref.\cite{PhysRevLett.108.070603} is unable to explain super-ballistic diffusion with the diffusion rate faster than cubic, so there is no correct explanation of the super-ballistic exponents with diffusion rate faster than cubic \cite{Nguyen2016}.
So a comprehensive interpretation of the super-ballistic exponents with diffusion rate faster than cubic is needed.

In this work, we studied the white noise effect on diffusion of the wave packet in 1D lattice. The white noise has different roles in these lattices, usually it suppresses the diffusion but in some cases it not only does not suppress the diffusion, but also will improve it. There is some especial case in which the system was localized and the white noise, delocalize the system and even drive the system to superballistic diffusion regime.
For the white noise effect on these tight-binding Hamiltonian, there is no numerical or analytic explanation nor interpretation available before, the suppression role of the white noise could be interpret via increasing the energy band mismatch between the sublattice and the rest uniform lattice and via the structure of the underlying eigenstates. The white noise changes quantum system's instantaneous eigenstate frequently, in \cite{PhysRevLett.vilar} the interplay between an externally added noise and the intrinsic noise of systems that relax fast toward a stationary state was analyzed theoretically. It was found that increasing the intensity of the external noise can reduce the total noise of the system. The output noise reduction is due to the fact that the system is driven into states with lower intrinsic noise, where the confinement effort is effectively greater. Here the intrinsic noise of system is the noise within the Hamiltonian eigenstates like what is in the Ref.\cite{PhysRevE.92.022150,PhysRevA.76.042127}  and the external noise is the induced white noise. \\ Addition of white noise enhances the diffusion rate in some cases and suppresses it in other cases. We think that in the cases that noise has the counter-intuitive aspect (enhancement), there is some stochastic resonance between the matter-wave states and the white noise, and so whenever this resonance occurs the noise enhances the diffusion otherwise it suppresses the diffusion. For sure the type of potential and it's intensity affects the occurrence of the stochastic resonance, the potential type, it's intensity and the noise intensity together have some correlation to the stochastic resonance, we have assessed many different cases, for sure the potential should be nested within a large enough perfect lattice, and it's intensity should be between certain values(with respect to the noise intensity, but it's structure could vary, here we use the triangular potential to present the superballistic diffusion, but even in fixed valued potential we have seen the localization to delocalization transition, yet it shows subballistic diffusion.

Acknowledgment. — The authors thank F.Shahbazi,  M. Amini And  V.Salari for their helpful comments. 
%
%

%
%

\bibliography{Ref}

\end{document}